\documentclass[aps,nofootinbib,twocolumn]{revtex4}

\newcommand{\be}{\begin{equation}}
\newcommand{\ee}{\end{equation}}
\newcommand{\bea}{\begin{eqnarray}}
\newcommand{\eea}{\end{eqnarray}}

\newcommand{\eq}[1]{Eq.~(\ref{eq:#1})}

\newcommand{\del}{\partial}
\newcommand{\tpi}{\tau_{\pi}}

\begin{document}

%%%%%%%   Title Page   %%%%%%%
%
%\rightline{KEK-TH-1133}
%\rightline{hep-ph/yymmnnn}
%
%\title{
%Comment on ``Viscous hydrodynamics relaxation time from AdS/CFT"
%}
%
%\author{Makoto Natsuume}
%\email{makoto.natsuume@kek.jp}
%\affiliation{Theory Division, Institute of Particle and Nuclear Studies, \\
%KEK, High Energy Accelerator Research Organization, Tsukuba, Ibaraki, 305-0801, Japan}
%
%\author{Takashi Okamura}
%\email{okamura@ksc.kwansei.ac.jp}
%\affiliation{Department of Physics, Kwansei Gakuin University,
%Sanda, Hyogo, 669-1337, Japan}
%
%\date{\today}
%
%\begin{abstract}
%Abstract here
%\end{abstract}
%
%\pacs{11.25.-w}
%
%\maketitle

%%%%%%%%%
\noindent
\textbf{Comment on ``Viscous hydrodynamics relaxation time from AdS/CFT correspondence"}
\bigskip
%%%%%%%%%
% KEK-TH-1210

Standard hydrodynamics does not satisfy causality, and the causal theory of hydrodynamics is known as ``causal hydrodynamics" \cite{Israel:1976tn,Israel:1979wp}. From an effective theory point of view, restoring causality forces one to consider higher orders in expansion. This means that causal hydrodynamics require a new set of transport coefficients in addition to ordinary transport coefficients such as shear viscosity $\eta$. One such coefficient is $\tpi$, the relaxation time for the shear viscous stress. Reference~\cite{Heller:2007qt} determines the coefficient of the ${\cal N}=4$ SYM from AdS/CFT correspondence. The purpose of this comment is to point out that the %likely 
value of $\tpi$ is 3 times larger than their result if one takes into account an additional term in the hydrodynamic equation.%
\footnote{Just before we submitted this comment, a number of interesting papers appeared \cite{Benincasa:2007tp,Baier:2007ix,Bhattacharyya:2007jc,Natsuume:2008iy}, which study the similar problem as ours. In particular, our point was made independently in Ref.~\cite{Benincasa:2007tp}.}
To make our point clear, the gravity computation done by Ref.~\cite{Heller:2007qt} itself remains valid [\eq{energy}], and the difference lies in the hydrodynamic interpretation.

Reference~\cite{Heller:2007qt} considers a ${\cal N}=4$ expanding plasma. For a boost-invariant plasma, which is often considered in the study of heavy-ion collisions, the following coordinate system is useful:
\be
ds^2 = -d\tau^2 + \tau^2dy^2 + dx_\perp^2~,
\ee
where $\tau$, $y$, and $x_\perp$ are proper time, rapidity, and the transverse coordinates, respectively. Basic equations in hydrodynamics are the conservation equation and the constitutive equation. For the expanding plasma, they are given by (See, {\it e.g.}, Ref.~\cite{Muronga:2001zk}.)
\bea
\del_\tau \epsilon &=& -\frac{\epsilon+p}{\tau} + \frac{\Phi}{\tau}~, 
   \label{eq:conserv} \\
\tpi \del_\tau \Phi &=& -\Phi + \frac{4\eta}{3\tau}
   \nonumber \\
&&-\frac{1}{2}\tpi \Phi 
\left\{ \frac{1}{\tau} + \frac{T}{\beta_2}\frac{d}{d\tau}\left(\frac{\beta_2}{T}\right) \right\}~,
   \label{eq:constitutive}
\eea
where $\epsilon$ is the energy density, $p$ is the pressure, $\Phi := -\tau^2\pi^{yy}$ is the dissipative part of the energy-momentum tensor, and $\beta_2 := \tpi/(2\eta)$. 

The last term in \eq{constitutive} needs some explanation because it plays an important role in our comment. The term is not included in the original Israel-Stewart theory (nor in Ref.~\cite{Heller:2007qt}), but it is mandatory to ensure the second law of thermodynamics. As far as we are aware, it has been first added by Muronga \cite{Muronga:2001zk}. On the other hand, the term is normally higher orders in the deviations from local equilibrium, so it is often neglected (See, {\it e.g.}, \cite{Heinz:2005bw}.) This may be the reason why it is not included in the original Israel-Stewart theory. However, such a naive power counting fails for a rapidly evolving system. In fact, one can easily check that the term is the same order as the other terms using \eq{energy} below.

The ${\cal N}=4$ SYM is conformal, so $\epsilon=3p$. Also, 
\be
\eta = C_\eta T^3, \qquad \tpi = \frac{C_\tau}{T}
\ee
from dimensional grounds. 
The aim of Ref.~\cite{Heller:2007qt} is to determine the constant $C_\tau$ (and $C_\eta$) from the gravity side.
The gravity computation yields the following energy density:
\be
\epsilon(\tau) =
  \frac{N_c^2}{2\pi^2} 
  \frac{1}{ \tau^{\frac{4}{3}} }
  \left\{ 1- \frac{\sqrt{2}}{ 3^{\frac{3}{4}} \tau^{\frac{2}{3}} } 
  + \frac{\sqrt{3}}{36} \frac{1+2\ln 2}{ \tau^{\frac{4}{3}} }+\cdots \right\}~.
\label{eq:energy}
\ee
We define the effective temperature $T$ as
%For the ${\cal N}=4$ SYM at strong coupling, $\epsilon$ and $T$ are related by 
\be
\epsilon = \frac{3}{8} \pi^2 N_c^2 T^4
\ee
which comes from the gravity computation for the ${\cal N}=4$ SYM at strong coupling. Note that temperature is $\tau$-dependent.
Substituting \eq{energy} into \eq{conserv} determines $\Phi$; then, $\Phi$ determines $C_\tau$ and $C_\eta$ from \eq{constitutive}. Ignoring the last term of \eq{constitutive}, one gets
\be
\tpi = \frac{1-\ln2}{6\pi T}~,
\ee
which is the value obtained in Ref.~\cite{Heller:2007qt}. However, taking the last term into account, one gets
\be
\tpi = \frac{1-\ln2}{2\pi T}~,
\label{eq:tau_pi}
\ee
which is 3 times larger.

This value of $\tpi$ is also supported from a computation in a different setting \cite{Natsuume:2007ty}. In order to avoid the confusion which comes from the power counting, it is best to consider a plasma whose deviations are small from equilibrium. In this case, the naive counting does work and we obtain the exactly the same value (\ref{eq:tau_pi}).

We thank Tetsufumi Hirano for discussions.  

\newpage

\bigskip

\noindent
Makoto Natsuume\\
$~~~$ Theory Division \\
$~~~$ Institute of Particle and Nuclear Studies \\
$~~~$ KEK \\%, High Energy Accelerator Research Organization\\ 
$~~~$ Tsukuba, Ibaraki, 305-0801, Japan\\

\noindent
Takashi Okamura\\
$~~~$ Department of Physics\\
$~~~$ Kwansei Gakuin University\\
$~~~$ Sanda, Hyogo, 669-1337, Japan\\

\end{document}